\documentclass[a4paper,11pt]{article}
\usepackage{pos}

\usepackage{slashed}  
\usepackage{multirow}

\allowdisplaybreaks[4]

\makeatletter
\newcommand{\figcaption}{\def\@captype{figure}\caption}
\newcommand{\tabcaption}{\def\@captype{table}\caption}

\newcommand{\Rmnum}[1]{\expandafter\@slowromancap\romannumeral #1@}

\def\hlinewd#1{%
  \noalign{\ifnum0=`}\fi\hrule \@height #1 \futurelet
   \reserved@a\@xhline}
\makeatother

\newcommand\dqq{\Big\langle \bar q q \Big\rangle}

\newcommand\dqGq{\Big\langle g_s \bar q \sigma G q \Big\rangle}

\newcommand\dGG{\Big\langle \alpha_{s} GG \Big\rangle}
\newcommand\dGGG{\Big\langle g_{s}^{3} G^{3} \Big\rangle}

\title{Light hybrid baryons in QCD sum rules}

\author*[a,b]{Wei Chen}
\author[a,c,d]{Qi-Nan Wang}
\author[a]{Ding-Kun Lian}
\author[e]{Hui-Min Yang}
\author[f]{Hua-Xing Chen}
\author[g]{J. Ho}
\author[h]{T. G. Steele}

\affiliation[a]{School of Physics, Sun Yat-Sen University, Guangzhou 510275, China}

\affiliation[b]{Southern Center for Nuclear-Science Theory (SCNT), Institute of Modern Physics,\\
Chinese Academy of Sciences, Huizhou 516000, Guangdong Province, China}

\affiliation[c]{Editorial Department of Journal, Bohai University, Jinzhou 121013, China}

\affiliation[d]{College of physical science and technology, Bohai University, Jinzhou 121013, China}

\affiliation[e]{School of Physics and Center of High Energy Physics, Peking University, Beijing 100871, China}

\affiliation[f]{School of Physics, Southeast University, Nanjing 210094, China}

\affiliation[g]{Department of Physics, Dordt University, Sioux Center, Iowa, 51250, USA}

\affiliation[h]{Department of Physics and Engineering Physics, University of Saskatchewan, Saskatoon, SK, S7N 5E2, Canada}

\emailAdd{chenwei29@mail.sysu.edu.cn}

\abstract{We have calculated the mass spectra of nucleon and delta hybrid baryons with both positive-parity and negative-parity by using the method of QCD sum rules. We predicte that the lowest-lying hybrid baryons are the negative-parity $N_{1/2^-}$ and $\Delta_{1/2^-}$ states, while the positive-parity ones are much heavier. We suggest to search for these light hybrid baryons via the $\chi_{cJ}/\Upsilon$ decay processes in the future.}

\FullConference{The 21st International Conference on Hadron Spectroscopy and Structure (HADRON2025)\\
 27 - 31 March, 2025\\
Osaka University, Japan\\}

\begin{document}
\maketitle

\section{Introduction}
The existence of hybrid states is one of the most distinctive predictions of quantum chromodynamics (QCD). The search for such exotic hadron states beyond conventional quark model (QM) has been ongoing for several decades in both experimental and theoretical aspects~\cite{Meyer:2015eta,Chen:2022asf}. 

To date, the most intriguing hybrid meson candidates are the isovector $\pi_{1}(1400)$~\cite{IHEP-Brussels-LosAlamos-AnnecyLAPP:1988iqi}, $\pi_{1}(1600)$~\cite{E852:2001ikk},  $\pi_{1}(2015)$~\cite{E852:2004gpn} and isoscalar $\eta_{1}(1855)$~\cite{BESIII:2022riz,BESIII:2022iwi} with exotic quantum numbers $J^{PC}=1^{-+}$, which have attracted extensive interest on their nature and properties~\cite{Meyer:2015eta,Chen:2022asf}. In contrast, the research on hybrid baryons are much less appealing, mainly due to the lack of exotic quantum numbers in the baryon sector. The traditional $qqq$ baryons in QM can achieve all $J^P$ quantum numbers, so that hybrid baryons can only appear as additional states within the baryon spectroscopy. In literatures, there have been some investigations on hybrid baryons in various theoretical approaches, such as the MIT bag model~\cite{Barnes:1982fj,Golowich:1982kx}, the constituent quark model~\cite{Burkert:2017djo}, the flux-tube model~\cite{Capstick:1999qq,Capstick:2002wm}, the relativistic quark model~\cite{Gerasyuta:2002hg}, the potential model~\cite{Cimino:2024bol}, the Lattice QCD~\cite{Dudek:2012ag,Khan:2020ahz} and the QCD sum rules~\cite{Martynenko:1991pc,Kisslinger:1995yw,Kisslinger:2003hk,Azizi:2017xyx,Zhao:2023imq,Wang:2024lnv,Yang:2025hzc}. 

In this talk, I shall introduce our recent calculations of the mass spectra of nucleon and delta hybrid baryons with both positive- and negative-parity in QCD sum rules. 

\section{QCD sum rules for light hybrid baryons}
The hybrid baryon interpolating currents with $J^P=1/2^{+}$ and $3/2^{+}$ are composed as the following~\cite{Wang:2024lnv}
\begin{equation}
 \begin{aligned}  \label{Eq:currents}
    J_{1}&=\varepsilon^{abc}g_s[(u^{aT}C\gamma ^{\mu}u^{b})\gamma^{\nu}\gamma^{5}(G_{\mu \nu }d)^{c}-(u^{aT}C\gamma ^{\mu}d^{b})\gamma^{\nu}\gamma^{5}(G_{\mu \nu }u)^{c}]\, ,\\
    J_{2}&=\varepsilon^{abc}g_s[(u^{aT}C\gamma ^{\mu}u^{b})\gamma^{\nu}\gamma^{5}(G_{\mu \nu }d)^{c}+2(u^{aT}C\gamma ^{\mu}d^{b})\gamma^{\nu}\gamma^{5}(G_{\mu \nu }u)^{c}]\, ,\\
     J_{3}&=\varepsilon^{abc}g_s[(u^{aT}C\sigma^{\mu\nu}u^{b})\gamma^{5}(G_{\mu \nu }d)^{c}-(u^{aT}C\sigma^{\mu\nu}d^{b})\gamma^{5}(G_{\mu \nu }u)^{c}]\, ,\\
    J_{4}&=\varepsilon^{abc}g_s[(u^{aT}C\sigma^{\mu\nu}u^{b})\gamma^{5}(G_{\mu \nu }d)^{c}+2(u^{aT}C\sigma^{\mu\nu}d^{b})\gamma^{5}(G_{\mu \nu }u)^{c}]\, ,\\
     J_{5}&=\varepsilon^{abc}g_s[(u^{aT}C\gamma ^{\mu}u^{b})\gamma^{\nu}(\tilde{G}_{\mu \nu }d)^{c}-(u^{aT}C\gamma ^{\mu}d^{b})\gamma^{\nu}(\tilde{G}_{\mu \nu }u)^{c}]\, ,\\
    J_{6}&=\varepsilon^{abc}g_s[(u^{aT}C\gamma ^{\mu}u^{b})\gamma^{\nu}(\tilde{G}_{\mu \nu }d)^{c}+2(u^{aT}C\gamma ^{\mu}d^{b})\gamma^{\nu}(\tilde{G}_{\mu \nu }u)^{c}]\, ,\\
       J_{7}&=\varepsilon^{abc}g_s[(u^{aT}C\sigma^{\mu\nu}u^{b})(\tilde{G}_{\mu \nu }d)^{c}-(u^{aT}C\sigma^{\mu\nu}d^{b})(\tilde{G}_{\mu \nu }u)^{c}]\, ,\\
    J_{8}&=\varepsilon^{abc}g_s[(u^{aT}C\sigma^{\mu\nu}u^{b})(\tilde{G}_{\mu \nu }d)^{c}+2(u^{aT}C\sigma^{\mu\nu}d^{b})(\tilde{G}_{\mu \nu }u)^{c}]\, ,\\
    J_{1\mu}&=\varepsilon^{abc}g_s[(u^{aT}C\gamma ^{\nu}u^{b})(G_{\mu \nu }d)^{c}-(u^{aT}C\gamma ^{\nu}d^{b})(G_{\mu \nu }u)^{c}]\, ,\\
    J_{2\mu}&=\varepsilon^{abc}g_s[(u^{aT}C\gamma ^{\nu}u^{b})(G_{\mu \nu }d)^{c}+2(u^{aT}C\gamma ^{\nu}d^{b})(G_{\mu \nu }u)^{c}]\, ,\\
        J_{3\mu}&=\varepsilon^{abc}g_s[(u^{aT}C\gamma ^{\nu}u^{b})\gamma^{5}(\tilde{G}_{\mu \nu }d)^{c}-(u^{aT}C\gamma ^{\nu}d^{b})\gamma^{5}(\tilde{G}_{\mu \nu }u)^{c}]\, ,\\
    J_{4\mu}&=\varepsilon^{abc}g_s[(u^{aT}C\gamma ^{\nu}u^{b})\gamma^{5}(\tilde{G}_{\mu \nu }d)^{c}+2(u^{aT}C\gamma ^{\nu}d^{b})\gamma^{5}(\tilde{G}_{\mu \nu }u)^{c}]\, ,\\
      \end{aligned}
  \end{equation}
in which the S-wave color-octet triquark operator $[qqq]_{8c}$ is adopted. The dual gluon field strength is defined as $\tilde{G}_{\mu \nu}=\frac{1}{2}\varepsilon_{\mu\nu\alpha\beta}G^{\alpha\beta}$. Among these interpolating currents, $J_{1,3,5,7,1\mu,3\mu }$ can couple to nucleon hybrid states, while $J_{2,4,6,8,2\mu,4\mu}$ couple to delta hybrids. They couple to both the positive- and negative-parity states via the different relations
\begin{equation}\label{Eq:coupling_positive}
\begin{split}
  \langle 0|J_{i}|H_{+}(p)\rangle&=f_{+} u(p),\\
    \langle 0|J_{i\mu}|H_{+}(p)\rangle&=f_{+} u_{\mu}(p),
    \end{split}
  \end{equation}
and
\begin{equation}\label{Eq:coupling_negative}
\begin{aligned}
  \langle 0|J_{i}|H_{-}(p)\rangle&=f_{-}\gamma_{5} u(p),\\
    \langle 0|J_{i\mu}|H_{-}(p)\rangle&=f_{-} \gamma_{5} u_{\mu}(p).
    \end{aligned}
  \end{equation}
where $u(p)$ and $u_{\mu }(p)$ are the Dirac spinor and Rarita-Schwinger vector, respectively, and $f_\pm$ is the corresponding coupling constant. 
The two-point correlation functions induced by currents $J(x)$ and $J_{\mu}(x)$ are
\begin{equation}\label{Eq:correlation}
  \begin{split}
\Pi(p^2)&=i\int  d^4x e^{ip\cdot x}\langle 0|T\left[J(x) \bar{J}(0)\right]|0\rangle=\Pi_{1/2}(p^2)\, ,\\
  \Pi_{\mu\nu }(p^2)&=i\int d^4x e^{ip\cdot x}\langle 0|T\left[J_{\mu}(x) \bar{J}_{\nu }(0)\right]|0\rangle=-g_{\mu\nu } \Pi_{3/2}(p^2)+\cdots\, ,
  \end{split}
  \end{equation}
where the invariant functions $\Pi_{1/2}(p^2)$ and $\Pi_{3/2}(p^2)$ contain both contributions from positive- and negative-parity states
\begin{align} \label{Eq:correlationParity}
  \Pi(p^2)=\slashed{p}\Pi_{\slashed{p}}+\Pi_{I}
=\left(\frac{\slashed{p}}{\sqrt{p^{2}}}+1\right)\Pi_{+}(p^2)+\left(\frac{\slashed{p}}{\sqrt{p^{2}}}-1\right)\Pi_{-}(p^2)\, .
   \end{align}
At the hadronic side, the invariant function can be described by the dispersion relation
\begin{equation}
\Pi(p^{2})=\frac{(p^{2})^{N}}{\pi} \int_{0}^{\infty} \frac{\operatorname{Im} \Pi(s)}{s^{N}\left(s-p^{2}-i \epsilon\right)} d s+\sum_{n=0}^{N-1} b_{n}(p^{2})^{n}\, .
\label{Cor-Spe}
\end{equation}
The spectral function can be defined in the ``narrow resonance'' approximation
\begin{align}\label{Eq:spectral}
  \rho(s) &\equiv \frac{\text{Im}\Pi (s)}{\pi   }
  =\slashed{p}\rho^{\slashed{p}}(s) + \rho^{I}(s)\\ \nonumber
  &=f_{+}^{\prime2}(\frac{\slashed{p}}{\sqrt{s}}+1)\delta (s-M_{+}^2)+f_{-}^{\prime2}(\frac{\slashed{p}}{\sqrt{s}}-1)\delta (s-M_{-}^2)+\cdots\,
\end{align}
where 
\begin{equation}
\begin{aligned}
\rho^{\slashed{p}}(s) & =\frac{f_{+}^{\prime2}}{\sqrt{s}}  \delta\left(s-M_{+}^{2}\right)+\frac{f_{-}^{\prime2}} {\sqrt{s}} \delta\left(s-M_{-}^{2}\right)+\cdots\, , \\
\rho^{I}(s) & =f_{+}^{\prime2}  \delta\left(s-M_{+}^{2}\right)-f_{-}^{\prime2}\delta\left(s-M_{-}^{2}\right)+\cdots\, ,
\end{aligned}
\end{equation}
with $f_{\pm}^{\prime2}=M_{\pm}f_{\pm}^{2}$. 
At the QCD side, the correlation functions and spectral functions can be evaluated via the operator product expansion (OPE) method, expressed as functions of various QCD condensates. 
Using the quark-hadron duality, one can establish the QCD sum rules after performing the Borel transform
\begin{equation}\label{Eq:PiFunction}
  \begin{split}
    \Pi_{\pm}(M_{B}^{2},s_{0})=\int _{(3m_q)^2}^{s_{0}} \rho_\pm(s) e^{-s/M_{B}^{2}}ds=2f_{\pm}^{\prime2}e^{-M_{\pm}^2/M_{B}^{2}} \, ,
     \end{split}
\end{equation}
where $\rho_{\pm}(s)=\text{Im}\Pi_{\pm}(s)/\pi=\sqrt{s}\rho^{\slashed{p}}(s)\pm\rho^{I}(s)$ contains information from positive/negative-parity state. 
The hadron mass can be extracted as
\begin{equation}
   \begin{aligned}
  M_{\pm}\left(M_B^{2},s_{0}\right) & =\sqrt{\frac{\frac{\partial}{\partial\left(-1 / M_B^2\right)} \Pi_{\pm}\left(M_B^{2},s_{0}\right)}{\Pi_{\pm}\left(M_B^{2},s_{0}\right)}} \,,
 \label{mass}
 \end{aligned}
\end{equation}
in which $M_B$ and $s_0$ are the Borel parameter and continuum threshold, respectively. 

\section{Numerical analyses}
The input parameters for numerical analyses are~\cite{ParticleDataGroup:2022pth,Narison:2011xe,Narison:2018dcr,Jamin:2002ev}
\begin{align}\label{parameters}
m_q&=0\, ,\nonumber \\
\alpha_s \left(\mu\right) & =\frac{4 \pi}{29/3 \ln \left(\mu^{2} / \Lambda_{\mathrm{QCD}}^2\right)}\, , \nonumber \\
\dqq & =-(0.24 \pm 0.01)^3 ~\mathrm{GeV}^3 \,,  \nonumber\\
\dqGq & =(0.8 \pm 0.2) \times \dqq ~\mathrm{GeV}^2 \,,  \\
\dGG & =(6.35 \pm 0.35) \times 10^{-2} ~\mathrm{GeV}^4 \,,  \nonumber\\
\dGGG & =(8.2 \pm 1.0) \times \dGG ~\mathrm{GeV}^2  \, ,\nonumber
\end{align}
where the renormalization scale $\mu=1 ~\mathrm{GeV}$ and $\Lambda_{QCD}=300 ~\mathrm{MeV}$ are adopted.

As shown in Eq.~(\ref{mass}), the extracted hadron mass is a function of two important parameters $s_{0}$ and $M_B^{2}$. To determine suitable working regions of them, we define the following two quantities to study the OPE convergence and pole contribution (PC)
\begin{align}
R_{D>6}&=\left|\frac{\Pi_{\pm}^{D>6}\left(M_B^{2}, \infty\right)}{\Pi_{\pm}^{t o t}\left(M_B^{2}, \infty\right)}\right| \, ,\label{Eqs:Conver} \\
\text{PC}&=\frac{\Pi_{\pm}\left(M_B^{2}, s_0\right)}{\Pi_{\pm}\left(M_B^{2}, \infty\right)}  \, .\label{Eqs:PC}
\end{align} 
where $\Pi_{\pm}^{D>6}\left(M_B^{2}, \infty\right)$ represents the contributions from $D>6$ condensates. 
In numerical analyses, the parameter working regions of $(s_0, M_B^{2})$ can be determined by the constrains of PC, $R_{D>6}$ and the stability of mass curves.

\begin{table}[t!]
\centering
  \caption{Numerical results for the nucleon and delta hybrid baryons.}
\renewcommand\arraystretch{1.3} 
  \setlength{\tabcolsep}{0.9em}{ 
    \begin{tabular}{ccccccc}
      \hline\hline  
       Current & State & $s_0[\mathrm{GeV}^2]$ & $M_B^2[\mathrm{GeV}^2]$ & Mass[GeV] & PC[\%] \\
      \hline 
$J_1$ & $N_{1/2^{-}}$ &  $11.0(\pm 5\%)$ & 2.75-2.87 & $3.01_{-0.06}^{+0.06}$  & 21.2  \\ 
      \cline { 2 - 7 }$J_2$ & $\Delta_{1/2^{-}}$& $16.5(\pm 5\%)$ & 2.53-4.25 &$3.64_{-0.08}^{+0.08}$  &34.7 \\
     \cline { 2 - 7 }       \multirow{2}{*}{$J_3$} & $N_{1/2^{-}}$ &  $6.5(\pm 5\%)$ & 1.59-2.05 & $2.28_{-0.06}^{+0.07}$  &26.8  \\  
   &    $N_{1/2^{+}}$ & $13.0(\pm 5\%)$ & 2.00-2.87  & $3.37_{-0.15}^{+0.28}$ &   29.3 \\    
      \cline { 2 - 7 }
$J_4$& $\Delta_{1/2^{-}}$ & $8.0(\pm 5\%)$ & 2.43-2.57 & $2.64_{-0.15}^{+0.39}$ & $10.9$ \\    
      \cline { 2 - 7 }
$J_5$ & $N_{1/2^{-}}$ &  $10.5(\pm 5\%)$ & 2.78-3.06 & $2.91_{-0.05}^{+0.05}$ & 22.6  \\  
      \cline { 2 - 7 }
$J_6$&  $\Delta_{1/2^{+}}$&  $20.0(\pm 5\%)$ & 3.09-3.32 &$4.03_{-0.11}^{+0.13}$ & 52.5 &  \\      
    \cline { 2 - 7 }      
$J_7$ & $N_{1/2^{-}}$ &  $14.5(\pm 5\%)$ & 2.21-3.44 & $3.48_{-0.11}^{+0.14}$ & 32.3  \\  
    \cline { 2 - 7 }
$J_8$ & $\Delta_{1/2^{-}}$ &  $12.5(\pm 5\%)$ & 2.02-3.56 & $3.15_{-0.06}^{+0.06}$  &36.1  \\  
   \cline { 2 - 7 }
   $J_{1 \mu}$ & $N_{3/2^{-}}$&  $17.5(\pm 5\%)$& 3.28-4.58 &$3.76_{-0.09}^{+0.10}$ & 29.6   \\
  \cline { 2 - 7 }
$J_{2 \mu}$ 
   & $\Delta_{3/2^{+}}$ &  $15.0(\pm 5\%)$ & 2.39-2.76 & $3.43_{-0.06}^{+0.06}$ & 55.6 & \\
  \cline { 2 - 7 }
  $J_{3 \mu}$ &$N_{3/2^{-}}$&  $12.0(\pm 5\%)$& 2.88-3.44 &$3.12_{-0.08}^{+0.08}$ & 24.9   \\
      \hline\hline  
  \end{tabular}
}
  \label{tab:results}
\end{table}

\begin{figure}[t!!]
\centering
\includegraphics[width=0.70\textwidth]{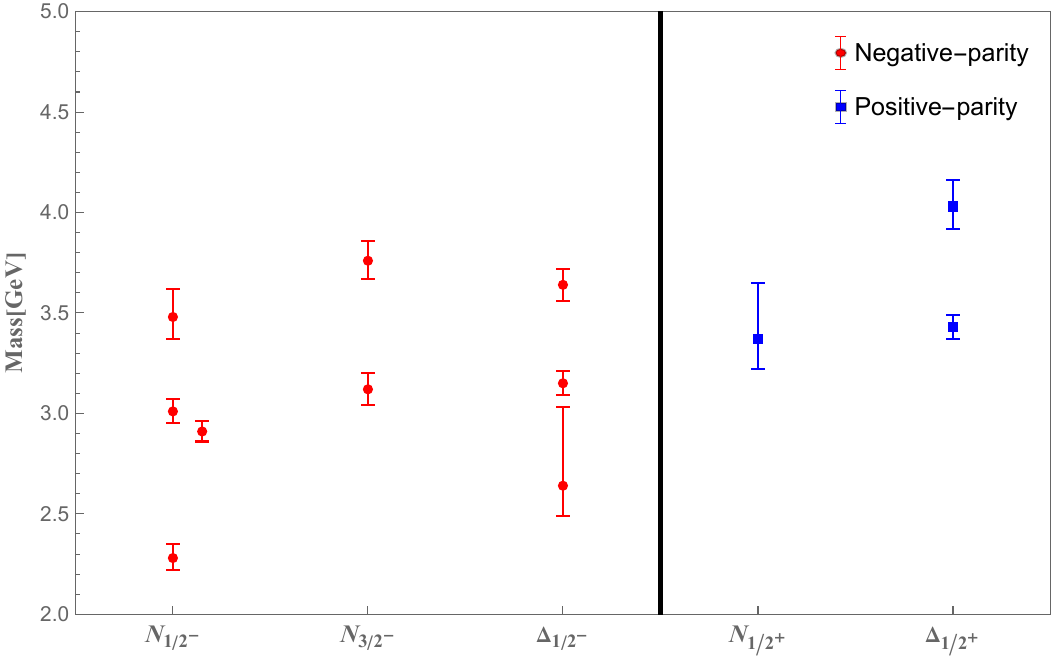}
\caption{Mass spectra for both negative-parity and positive-parity hybrid baryons.}
\label{fig:mass}
\end{figure}

After performing numerical analyses for all interpolating currents in Eq.~(\ref{Eq:currents}), we establish stable mass sum rules for the positive-parity $N_{1/2^+}, \Delta_{3/2^+}, \Delta_{1/2^+}$ 
and negative-parity $N_{1/2^-}, N_{3/2^-}, \Delta_{1/2^-}$ channels. We collect the numerical results for these hybrid baryons in Table~\ref{tab:results}. 
As depicted in Fig.~\ref{fig:mass}, the lightest hybrid baryons are predicted to be the negative-parity $N_{1/2^-}$ around 2.28 GeV and $\Delta_{1/2^-}$ state around 2.64 GeV. All positive-parity hybrid baryons are predicted to be higher than 3 GeV. Moreover, there is no stable nucleon hybrid baryon $N_{3/2^+}$ in our results.

\section{Summary} 
In this talk, I have introduced our calculations of mass spectra for the light nucleon and delta hybrid baryons in parity-projected QCD sum rule method. 
We predict that the lightest hybrid baryons are the negative-parity $N_{1/2^-}$ and $\Delta_{1/2^-}$, while the positive-parity states are much heavier. 
These hybrid baryons may primarily decay into the conventional baryon-meson final states, which are expected to be produced via the gluon-rich processes in the BESIII and BelleII experiments. 
\\

{\bf \noindent ACKNOWLEDGMENTS:}
This work is supported by the National Natural Science Foundation of China under Grant No. 12305147, No. 12175318 and No. 12075019, the Natural Science Foundation of Guangdong Province of China under Grants No. 2022A1515011922. TGS is grateful for research funding from the Natural Sciences \& Engineering Research Council of Canada (NSERC). JH was supported by the Mitacs Globalink Research Award and is grateful for the support and hospitality of SYSU.


\begin{thebibliography}{99}

\bibitem{Meyer:2015eta}
C.~Meyer and E.~Swanson, Prog. Part. Nucl. Phys. {\bf 82}, 21 (2015)

\bibitem{Chen:2022asf}
H.-X. Chen, W.~Chen, X.~Liu, Y.-R. Liu, and S.-L. Zhu, Rept. Prog. Phys. {\bf
  86}, 026201 (2023)
  
\bibitem{IHEP-Brussels-LosAlamos-AnnecyLAPP:1988iqi}
D.~Alde et~al., Phys. Lett. B {\bf 205}, 397 (1988)

\bibitem{E852:2001ikk}
E.~I. Ivanov et~al., Phys. Rev. Lett. {\bf 86}, 3977 (2001)

\bibitem{E852:2004gpn}
J.~Kuhn et~al., Phys. Lett. B {\bf 595}, 109 (2004)

\bibitem{BESIII:2022riz}
M.~Ablikim et~al., Phys. Rev. Lett. {\bf 129}, 192002 (2022)

\bibitem{BESIII:2022iwi}
M.~Ablikim et~al., Phys. Rev. D {\bf 106}, 072012 (2022)
  
\bibitem{Barnes:1982fj}
T.~Barnes and F.~E. Close, Phys. Lett. B {\bf 123}, 89 (1983)

\bibitem{Golowich:1982kx}
E.~Golowich, E.~Haqq, and G.~Karl, Phys. Rev. D {\bf 28}, 160 (1983)

\bibitem{Burkert:2017djo}
V.~D. Burkert and C.~D. Roberts, Rev. Mod. Phys. {\bf 91}, 011003 (2019)

\bibitem{Capstick:1999qq}
S.~Capstick and P.~R. Page, Phys. Rev. D {\bf 60}, 111501 (1999)

\bibitem{Capstick:2002wm}
S.~Capstick and P.~R. Page, Phys. Rev. C {\bf 66}, 065204 (2002)

\bibitem{Gerasyuta:2002hg}
S.~M. Gerasyuta and V.~I. Kochkin, Phys. Rev. D {\bf 66}, 116001 (2002)

\bibitem{Cimino:2024bol}
L.~Cimino, C.~T. Willemyns, and C.~Semay, Phys. Rev. D {\bf 110}, 034032 (2024)

\bibitem{Dudek:2012ag}
J.~J. Dudek and R.~G. Edwards, Phys. Rev. D {\bf 85}, 054016 (2012)

\bibitem{Khan:2020ahz}
T.~Khan, D.~Richards, and F.~Winter, Phys. Rev. D {\bf 104}, 034503 (2021)

\bibitem{Martynenko:1991pc}
A.~P.~Martynenko,
Sov. J. Nucl. Phys. \textbf{54} (1991), 488-490

\bibitem{Kisslinger:1995yw}
L.~S. Kisslinger and Z.~P. Li, Phys. Rev. D {\bf 51}, R5986 (1995)

\bibitem{Kisslinger:2003hk}
L.~S. Kisslinger, Phys. Rev. D {\bf 69}, 054015 (2004)

\bibitem{Azizi:2017xyx}
K.~Azizi, B.~Barsbay, and H.~Sundu, Eur. Phys. J. Plus {\bf 133}, 121 (2018)

\bibitem{Zhao:2023imq}
Y.-C. Zhao, C.-M. Tang, and L.~Tang, Eur. Phys. J. C {\bf 83}, 654 (2023)

\bibitem{Wang:2024lnv}
Q.~N.~Wang, D.~K.~Lian, W.~Chen, H.~M.~Yang, H.~X.~Chen, J.~Ho and T.~G.~Steele,
[arXiv:2412.14878 [hep-ph]].

\bibitem{Yang:2025hzc}
H.~M.~Yang, X.~Luo, H.~X.~Chen and W.~Chen,
[arXiv:2505.18717 [hep-ph]].

\bibitem{ParticleDataGroup:2022pth}
R.~L. Workman et~al., PTEP {\bf 2022}, 083C01 (2022)

\bibitem{Narison:2011xe}
S.~Narison, Phys. Lett. B {\bf 706}, 412 (2012)

\bibitem{Narison:2018dcr}
S.~Narison, Int. J. Mod. Phys. A {\bf 33}, 1850045 (2018)

\bibitem{Jamin:2002ev}
M.~Jamin, Phys. Lett. B {\bf 538}, 71 (2002)


\end{thebibliography}
\end{document}